\documentclass[aps,amsmath,amssymb,amsfonts,prd,nofootinbib,reprint,nobibnotes]{revtex4-2}

\pdfoutput=1 

\usepackage{graphicx}
\usepackage{hyperref}
\usepackage{caption,subcaption} 
\usepackage[section]{placeins}

\newcommand{\comment}[1]{}
\newcommand{\beq}[1]{\begin{equation}\label{#1}}
\newcommand{\eeq}{\end{equation}}

\newcommand{\C}{\mathcal{C}}
\newcommand{\B}{\mathcal{B}}
\newcommand{\del}{\partial}
\newcommand{\Del}{\nabla}

\newcommand{\N}{\mathcal{N}}
\newcommand{\bea}{\begin{eqnarray}}
\newcommand{\eea}{\end{eqnarray}}
\renewcommand{\t}{\tilde}

\renewcommand{\ap}{\alpha'}
\newcommand{\w}{\wedge}
\newcommand{\OmT}{\Omega_{T}}
\newcommand{\F}{\mathcal{F}}
\newcommand{\csch}{\textnormal{csch}}
\newcommand{\G}{\mathcal{G}}

\begin{document}

\title{Holographic complexity of the Klebanov-Strassler background}

\author{Andrew R.~Frey}
\email{a.frey@uwinnipeg.ca}
\affiliation{Department of Physics and Winnipeg Institute for
Theoretical Physics, University of Winnipeg\\
515 Portage Avenue, Winnipeg, Manitoba R3B 2E9, Canada}

\author{Michael P.~Grehan}
\email{michael.grehan@mail.utoronto.ca}
\affiliation{Department of Physics, University of Toronto\\
60 St. George St., Toronto, Ontario M5S 1A7, Canada}
\affiliation{Canadian Institute for Theoretical Astrophysics\\ 
60 St. George St., Toronto, Ontario M5S 3H8, Canada}

\author{Prakriti Singh}
\email{psingh39@syr.edu}
\affiliation{Department of Physics, Indian Institute of Technology Kanpur\\
Kanpur 208016, India}
\altaffiliation[Current Address:~]{Department of Physics,
Syracuse University, Syracuse, NY 13244, USA}

\begin{abstract}We study the complexity of the gravity dual to the confining 
$SU(N)\times SU(N+M)$ Klebanov-Strassler gauge theory, which is an 
important test case for holographic complexity in higher-dimensional and
nonconformal gauge-gravity dualities. We emphasize the
dependence of the complexity on parameters of the gauge theory, finding a
common behavior with confinement scale for several complexity functionals.
We also analyze how the complexity diverges with the UV cutoff, which is
more complicated than in AdS backgrounds because the theory is nonconformal.
Our results may provide new perspectives on questions in the holographic
complexity program as well as a starting point for further studies of
complexity in general gauge-gravity dualities.
\end{abstract}

\maketitle

\section{Introduction}

In quantum information theory, (circuit) complexity is a measure of the
distance of a quantum state to some reference state; holographic complexity
for a gauge theory is the dual of complexity on the gravity side of a 
gauge-gravity duality such as the AdS/CFT correspondence (see
\cite{arXiv:2110.14672} for references and a review).
However, the definition of complexity in quantum mechanics or quantum 
field theory is not unique
because there are many possible definitions of distance in the space of
states \cite{quant-ph/0502070,quant-ph/0603161,quant-ph/0701004,arXiv:1707.08570,arXiv:1707.08582,arXiv:1801.07620,arXiv:1803.10638}. 
Similarly, there is an infinite family of functionals that evaluate 
holographic complexity. In general, 
holographic complexity is a codimension-1 or -0 functional evaluated on a 
surface or region extending from a slice $\Sigma$ of the conformal boundary
space at a cutoff value of the holographic radius (which we will call the
boundary for simplicity). Via duality, this corresponds to the gauge
theory's complexity on the given time slice $\Sigma$. 

There are several
properties we may expect from a general complexity functional, including
late-time linear growth and the switchback effect in black hole spacetimes
\cite{arXiv:2111.02429,arXiv:2210.09647} or its value as the strongest
possible divergence in pure AdS spacetime \cite{arXiv:2307.08229}.
And the most commonly studied holographic complexity functionals, including
the CV ``complexity=volume'' proposal \cite{arXiv:1403.5695,arXiv:1406.2678},
the CV2.0 proposal \cite{arXiv:1610.02038}, 
and the CA ``complexity=action'' proposal 
\cite{arXiv:1509.07876,arXiv:1512.04993},
satisfy these properties. However, with the exception of 
\cite{arXiv:2109.06883,arXiv:2111.14897,frey}, discussion of holographic
complexity has centered on asymptotically AdS spacetimes without the
extra dimensions in the fully string theoretic gauge-gravity duality.

Here, we examine the holographic complexity of the Klebanov-Strassler (KS)
gauge theory $\mathbb{Z}_2$ symmetric ground state. 
The KS gauge theory \cite{hep-th/0007191} is an 
$\N=1$ theory with gauge group $SU(N+M)\times SU(N)$ with four chiral 
superfields, two of which transform in the bifundamental and two in its 
conjugate. This theory is holographic, and \cite{hep-th/0007191} proposed that 
the gravity dual is a ten-dimensional 
warped product of Minkowski spacetime with a deformed conifold along 
with 3-form and 5-form flux (the KS background 
in IIB supergravity). Since the KS background
is static, there is an unambiguous time-independent complexity, 
and we can choose $\Sigma$ to be the $t=0$ slice of the boundary. 

The gauge theory has several important properties, one of which is 
that it is confining in the infrared, 
which corresponds to the smooth and horizonless nature of the gravity dual, 
similar to AdS soliton backgrounds \cite{hep-th/9808079,hep-th/9803131}. 
As with complexity of the soliton 
backgrounds \cite{arXiv:1712.03732,arXiv:2307.08229}, how the complexity of
the KS background behaves as a function of the confinement scale 
(given by the conifold deformation parameter $\epsilon$) is an important 
question. We find a universal scaling with $\epsilon^2$, which is equivalent
to the cube of the confinement scale, when we write the UV cutoff
in terms of the dimensionless radial coordinate.
We also investigate the dependence of complexity on the gauge group rank
parameter $M$ (a flux quantum number in the KS background) and gauge
coupling as given by $g_s$.

In addition to confinement in the infrared, the gauge theory undergoes 
a series of dualities (a ``duality cascade'') to an $SU(M)$ gauge group.
Conversely, as the renormalization scale moves toward the ultraviolet,
the rank of the gauge group continues to increase. 
That is, it is also nonconformal at high 
energies, so the holographic complexity has a different divergence structure
as a function of cutoff radius
for asymptotically KS backgrounds as opposed to asymptotically AdS backgrounds.
However, at large radius, the KS background is geometrically similar to 
AdS with a slowly varying curvature radius. As a result, well-known results
on holographic complexity in AdS give us an expectation for the leading
divergence of complexity in KS. We find the leading divergences for
various holographic complexity proposals as well as a numerical evaluation
of the full holographic complexity for comparison. These results should be
useful as a reference for future studies of black brane (thermal) solutions
in KS \cite{arXiv:0706.1768,arXiv:1809.08484,arXiv:2103.15188}, 
particularly in terms of defining a version of
the complexity of formation 
\cite{arXiv:1509.07876,arXiv:1512.04993,arXiv:1610.08063} 
for asymptotically KS spacetimes.

Further, the KS background actually lies on a one-complex-parameter
moduli space of the field theory 
parametrized by the expectation value of baryonic operators 
(the ``baryonic branch'') at a point of $\mathbb{Z}_2$ symmetry;
\cite{hep-th/0405282,hep-th/0412187} described the gravity dual to the
baryonic branch (perturbatively and numerically respectively). 
We show that the KS background is a local extremum of the
complexity functionals that we consider along the baryonic branch. 


We begin with a review of the KS
background in Sec.~\ref{s:ksreview} below. Then we discuss the behavior
of several holographic complexity functionals in subsequent sections,
including a revised CA complexity and the ``complexity=flux'' proposals
of \cite{frey}.
In each case, we review the definition of the functional along with comments
on their use in 10D gauge-gravity dualities. Then we give their
scaling with the physical parameters of the gauge theory and determine
the divergence structure analytically in the cutoff radius. We also
give a numerical evaluation of each functional as a function of cutoff
radius. In Sec.~\ref{s:baryonic}, 
we argue that the $\mathbb{Z}_2$ symmetric KS solution 
extremizes the complexity along the baryonic branch of the gauge theory.
We conclude with a discussion of their implications
and connections to other research in \ref{s:discussion}.


\section{Review of KS background}\label{s:ksreview}

Here we give a brief review of the KS background \cite{hep-th/0007191},
as influenced by the discussion of \cite{hep-th/0108101}.

The KS background is a solution of the 10D type IIB
supergravity with metric
\beq{ksmetric}
ds^2=h(\tau)^{-1/2} \eta_{\mu\nu} dx^\mu dx^\nu + h(\tau)^{1/2} d\t s^2 ,\eeq
where $x^\mu$ are the boundary coordinates $t,\vec x$ and
\bea
d\t s^2 &=& \frac 12 \epsilon^{4/3} K(\tau) \left[ \frac{1}{3K(\tau)^3}
\left(d\tau^2+(g_5)^2\right)\right.\nonumber\\
&&\left. +\sinh^2\left(\frac\tau 2\right) \left(
(g_1)^2+(g_2)^2\right)\right.\nonumber\\
&&\left.\vphantom{\frac{1}{K^3}} +\cosh^2\left(\frac\tau 2\right) \left( (g_3)^2
+(g_4)^2\right)\right]\label{conifold} \eea
is the metric of the deformed conifold for 
$K(\tau)=(\sinh(2\tau)-2\tau)^{1/3}/(2^{1/3}\sinh(\tau))$ and deformation
modulus $\epsilon$ (which sets the confinement scale in the gauge theory). 
$\tau$ is the
radial (holographic energy) direction, and the angular basis forms are
\bea
&&g_1=\frac{e_1-e_3}{\sqrt 2} ,\quad g_2=\frac{e_2-e_4}{\sqrt 2} ,\quad
g_3=\frac{e_1+e_3}{\sqrt 2} ,\nonumber\\
&& g_4=\frac{e_2+e_4}{\sqrt 2} ,\quad g_5=e_5
\label{gbasis}\eea
with
\bea e_1&=&-\sin\theta_1 d\phi_1 ,\quad e_2=d\theta_1 ,\nonumber\\  
e_3&=&\cos\psi\sin\theta_2 d\phi_2-\sin\psi d\theta_2 ,\nonumber\\ 
e_4&=&\sin\psi\sin\theta_2 d\phi_2+\cos\psi d\theta_2 ,\nonumber\\ 
e_5&=&d\psi +\cos\theta_1 d\phi_1 +\cos\theta_2 d\phi_2 .\label{ebasis}\eea
Since KS is translation invariant
along the boundary, any measure of complexity is formally infinite (proportional
to boundary volume), so we will actually measure the density as measured
with respect to the coordinates $\vec x$.
We denote complexity densities as $\C$. Further, the $\C$ functionals 
we evaluate below will all contain a common integral of 
$\sin\theta_1\sin\theta_2$ over the angular coordinates, the angular
volume of $T^{1,1}$, which we will denote $\OmT$.

While the deformed conifold is Ricci-flat, the warp factor $h(\tau)$ results
from supergravity 3-form fluxes
\bea
F_3&=& \left(\frac{M\ap}{2}\right)\left[\vphantom{\frac 12} F g_5\w g_1\w g_2
+(1-F)g_5\w g_3\w g_4\right.\nonumber\\
&&\left. \vphantom{\frac 12}+ F' d\tau \w (g_1\w g_3+g_2\w g_4) \right] ,
\label{3forms}\\
H_3 &=& \left(\frac{g_s M\ap}{2}\right)\left[ d\tau\w
(f'_-\, g_1\w g_2+f'_+\, g_3\w g_4)\vphantom{\frac 12}\right.\nonumber\\
&&\left. +\frac 12 (f_+-f_-)^2 
g_5\w(g_1\w g_3+g_2\w g_4)\right] ,\nonumber
\eea
where $g_s$ and $\ap$ are the string coupling and Regge parameter, 
respectively, 
(as standard convention) and $M$ is the flux quantum number, which determines
the gauge group in the IR. The 3-forms satisfy the duality relation
$g_s F_3=-\t\star H_3$. The functions $f_\pm$ and $F$ are given by
\beq{miscfuncs} f_\pm(\tau)=\left(\frac{\tau\coth\tau - 1}{2\sinh\tau}\right)
(\cosh\tau \pm 1) ,\ F(\tau)=\frac{\sinh\tau-\tau}{2\sinh\tau} .\eeq
These source a 5-form flux $\t F_5=(1+\star)\F_5$ with
\beq{5form}
\F_5 = \left(\frac{g_s M^2\ap^2}{4}\right) l g_1\w g_2\w g_3\w g_4\w g_5 ,
\ l=f_+F+f_-(1-F)\eeq
and warp factor 
\bea
h(\tau)&=&\left(\frac{2^{1/3}g_s M\ap}{\epsilon^{4/3}}\right)^2  I(\tau) ,
\label{warping}\\
I(\tau) &=& \int_\tau^\infty dx \left(\frac{x\coth x-1}{\sinh^2 x}\right)
\left(\vphantom{\frac 12}\sinh(2x)-2x\right)^{1/3} .\nonumber\eea

At large $\tau$, the integral 
\beq{warping2}
I(\tau)\to (3/2^{1/3})\exp(-4\tau/3)(\tau-1/4).\eeq
With the coordinate transformation 
\beq{rcoord}r^2=(3\epsilon^{4/3}/2^{5/3})\exp(2\tau/3),\eeq
the large-radius metric becomes
\beq{ktmetric}
ds^2 = \frac{r^2}{L(r)^2}\eta_{\mu\nu} dx^\mu dx^\nu +\frac{L(r)^2}{r^2} dr^2
+L(r)^2 d\Omega_{T^{1,1}}^2 ,
\eeq
as first found in \cite{hep-th/0002159}. Here, 
$d\Omega_{T^{1,1}}^2$ is the metric on $T^{1,1}$ (an $S^1$ bundle over 
$S^2\times S^2$) and $L(r)$ is a radially dependent curvature scale given by
\beq{varyL}
L(r) = \left(\frac{9g_s M\ap}{2\sqrt 2}\right)^{1/2} \left(\ln\left(
\frac{2^{5/6} r}{\sqrt 3\epsilon^{2/3}}\right)\right)^{1/4} .\eeq
Since (\ref{ktmetric}) is AdS$_5\times T^{1,1}$ with a slowly varying 
curvature, we expect the leading divergence of complexity for KS to be 
that of AdS with an additional logarithmic radial dependence.

The KS solution falls in a continuous family of supersymmetric 
backgrounds dual to a
moduli space of the KS gauge theory known as the baryonic branch of the
theory. The KS solution has a $\mathbb{Z}_2$ symmetry given by the exchange
$(\theta_1,\phi_1)\leftrightarrow (\theta_2,\phi_2)$ in the $T^{1,1}$ 
angular directions. More general baryonic branch solutions follow an 
interpolating ansatz given by \cite{Papadopoulos:2000gj}, which breaks the 
$\mathbb{Z}_2$ symmetry; this geometry is sometimes known as a 
``resolved warped deformed conifold'' \cite{hep-th/0405282}. 
The gravity background for the full baryonic branch (as a solution of the
supersymmetry conditions) appears in numerical form in \cite{hep-th/0412187}.
We will describe perturbations of the KS solution along the baryonic
branch (first found in \cite{hep-th/0405282}) in Sec.~\ref{s:baryonic}.

\comment{
DISCUSSION OF BARYONIC BRANCH
\begin{itemize}\item``deformed resolved conifold'' and basis forms
\item warped geometry/flux at and near KS point
\item \cite{hep-th/0412187} Grana et al baryonic branch
\cite{hep-th/0405282} GHK baryonic branch
\end{itemize}
}

\section{Volume complexity functionals}

We first discuss the two common volume complexity functionals using volumes
defined in the ten-dimensional spacetime.

\subsection{CV volume complexity}\label{s:cv}

\subsubsection{Review}\label{ss:cvreview}

Possibly the simplest proposal for holographic complexity is the CV,
or ``complexity=volume'' proposal \cite{arXiv:1403.5695,arXiv:1406.2678}. 
The CV complexity is given by 
$C_V=V/G\ell$, where $V$ is the volume of the slice $\B$ 
(codimension-1) of the bulk spacetime
that maximizes volume subject to $\del\B=\Sigma$. $G$ is the 
Newton constant, and $\ell$ is an arbitrary length scale included to make
the complexity dimensionless; in asymptotically AdS backgrounds, $\ell$ is
often chosen equal to the AdS scale. In a ten-dimensional holographic
background like KS, there are different ways to define the maximization 
procedure to find $V$ (see \cite{arXiv:2109.06883,arXiv:2111.14897,frey}).
We will choose $\B$ as a nine-dimensional spatial slice of the 
full ten-dimensional
spacetime and $G$ as the ten-dimensional Newton constant. For spacetimes
that factorize into a holographic spacetime (such as asymptotically AdS
spacetime) and a compact space, this automatically equals the usual CV
complexity in the holographic spacetime with the same length $\ell$.
For example, consider AdS$_5\times S^5$ with Poincar\'e coordinates,
\beq{ads5s5}
ds^2 = \frac{r^2}{L^2}\eta_{\mu\nu} dx^\mu dx^\nu +\frac{L^2}{r^2} dr^2
+L^2 d\Omega_5^2\ .\eeq
With a radial cutoff at $r_m$, this complexity is 
\beq{ads5cmplx} \C_V = \frac{L^5 \Omega_5}{3 G_{10} \ell}\frac{r_m^3}{L^2}
= \frac{1}{3 G_{5}}\left(\frac L \ell\right)\left(\frac{r_m}{L}\right)^3
 ,\eeq
where $G_{10}$ and $G_5$ are the Newton constants in 10D and 5D, respectively,
and $\Omega_5$ is the volume of the unit $S^5$.
This result will be useful for understanding the divergence structure of
complexity in KS. Note that the complexity scales differently with the
AdS scale depending on whether we treat $G_{10}$ or $G_5$ as fixed and that 
it agrees with the CV complexity of AdS$_5$ for $G_5$ fixed (because we
have chosen a product spacetime background). In this paper, we will take
a definition with $G_{10}$ fixed.

\subsubsection{Parameter dependence}\label{s:cvscaling}

In the KS background, the CV complexity is simply given by the 
volume of the $t=0$ slice. We therefore need the 9D spatial volume
element, which is a major ingredient of all complexity functionals. 
Including the determinant of the deformed conifold and the power
$h(\tau)^{3/4}$ of the warp factor, this spatial volume is the integral of
\beq{velement}
\frac{\sqrt 2(g_s M\ap)^{3/2}\epsilon^2}{96} \left[\sin\theta_1\sin\theta_2\right]
\left[I(\tau)^{3/4}\sinh^2\tau\right] .\eeq
In total, the CV complexity density is
$\C_V=\sqrt 2 (g_s M\ap)^{3/2}\epsilon^2\OmT J_V/96G_{10}\ell$, where $J_V$ is
the $\tau$ integral of the factor in the last square brackets.

The fact that $\C_V\propto\epsilon^2$ is notable. Since $\epsilon$ has
dimension of length$^{3/2}$, this is the same scaling of complexity with
confinement scale as found for AdS soliton solutions in
\cite{arXiv:1712.03732,arXiv:2307.08229}. Furthermore, we notice that 
$\C_V\propto M^{3/2}$, in comparison to the AdS$_5\times S^5$ case in which
$\C_V\propto N^{3/4}$, with $N$ the D3-brane number. This is because the
effective D3-brane charge at large distances in KS scales as $M^2$; similar
comments apply to the scaling with $g_s$.

\subsubsection{Divergence structure}\label{s:cvdivergence}

\begin{figure}[t]\centering
\includegraphics[width=\columnwidth]{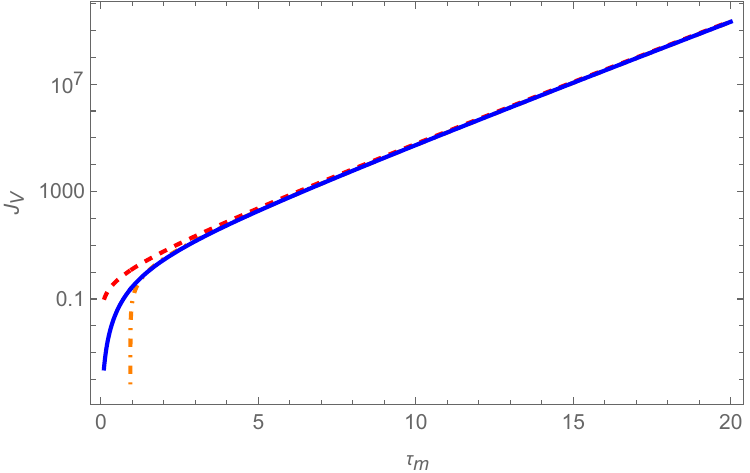}
\caption{\label{f:cvplot} The dimensionless CV complexity $J_V$ as a function
of cutoff $\tau_m$. Compares numerical calculation [solid blue, 
Eq.~\eqref{cvJ}], leading divergence [dashed red, \eqref{cvJleading}], 
and leading 
plus next-to-leading divergences [dot-dashed orange, \eqref{cvJleading}].}
\end{figure}

The overall value of the complexity depends on the radial integral 
\beq{cvJ}
J_V(\tau_m) \equiv \int_0^{\tau_m}d\tau\, I(\tau)^{3/4}\sinh^2\tau\eeq
from the origin to a maximal value $\tau_m$, acting as a UV cutoff. 
While the value for any specific cutoff is not physically interesting, 
the dependence on the cutoff scale is. It provides a glimpse at the 
high-energy degrees of freedom of the gauge theory, and it is a necessary
component of any definition of complexity of formation in asymptotically KS
backgrounds. To evaluate $J_V(\tau_m)$, we first evaluate $I(\tau)$,
the functional form of the warp factor, numerically, and then evaluate
$J_V(\tau_m)$ numerically. Figure \ref{f:cvplot} shows $J_V(\tau_m)$; at 
large $\tau_m$ it is nearly exponential.

We would also like to know the form of the leading divergence. Using the
asymptotic behavior of $I(\tau)$ from above (\ref{ktmetric}), the divergent
behavior is 
\beq{cvJdivergent}
J_V(\tau_m) \sim \left(\frac 38\right)^{3/4} \int^{\tau_m} d\tau\left(\tau
-\frac 14\right)^{3/4} e^\tau ,\eeq
with other contributions exponentially suppressed and convergent.\footnote{We
take $\sim$ to indicate that the divergences are identical, while $\approx$
will indicate an approximation that improves at large argument.} 
Note that the choice of any constant lower limit does not affect the functional
form of the divergence. If we choose the lower limit to be $1/4$, then
\bea
J_V(\tau_m)&\!\!\sim\!\!& \left(\frac 38\right)^{3/4}
e^{1/4}\gamma_1\left(\frac 74,\tau_m-\frac 14 \right) \label{cvJdivergent2}\\
&=&\left(\frac 38\right)^{3/4}
e^{1/4}\left(\tau_m-\frac 14\right)^{7/4}\Gamma\left(\frac 74\right)\nonumber\\
&&\times\gamma^*\left(\frac 74,\frac 14-\tau_m\right) ,\nonumber
\eea
where 
\beq{tricomi}
\gamma_1(a,x) = \int_0^x dt\, t^{a-1}e^t\ ,\
\gamma^*(a,z)=\frac{1}{\Gamma(a)}\int_0^1 dt\, t^{a-1}e^{-zt}
\eeq
are incomplete gamma functions following Tricomi \cite{tricomi1,tricomi2}
and B\"ohmer \cite{bohmer} [with $\text{Re}(a)>0$ for $\gamma_1$ while
$\gamma^*$ is entire in both arguments]. The $\gamma^*$ function has an 
expression in terms of confluent hypergeometric functions.

The gamma functions above give the divergence as an exponential of $\tau_m$
times a series with decreasing powers of $\tau_m$. 
To illustrate the behavior of the divergence, consider the leading 
and next-to-leading terms
\bea
J_V(\tau_m) &\approx& \left(\frac 38\right)^{3/4} \int^{\tau_m} d\tau\,\tau^{3/4}
\left(1-\frac{3}{16\tau}\right) e^\tau\nonumber \\
&\approx&
\left(\frac 38\right)^{3/4} \tau_m^{3/4} \left(1-\frac{15}{16\tau_m}\right)
e^{\tau_m} .\label{cvJleading}\eea
Figure \ref{f:cvplot} compares the numerically calculated $J_V(\tau_m)$ 
(shown in solid blue) to the leading divergence (dashed red) and first two 
divergences (dot-dashed orange). As the expansion (\ref{cvJleading}) 
suggests, including the leading and next-to-leading terms gives an approximation
valid to within a percent for $\tau_m\gtrsim 6$, while the leading term 
is accurate to within a percent for $\tau_m\gtrsim 90$.
In the radial variable $r$, the leading divergence becomes
\beq{cvleadingr}
\C_V=\frac{\OmT}{(18)^2G_{10}\ell} L(r_m)^3 r_m^3 ,
\eeq
which is parametrically identical to the complexity of AdS$_5\times S^5$
with AdS length chosen equal to the curvature scale of KS at $r_m$. Since
a common picture of KS at large radius is that of an AdS throat with 
slowly varying curvature, this result is very intuitive. It serves to 
demonstrate that the UV degrees of freedom dominate the complexity.


\subsection{CV2.0 volume complexity}\label{s:cv20}

\subsubsection{Review}\label{s:cv20review}

Another purely geometric holographic complexity functional, known as CV2.0,
is the volume of the Wheeler-DeWitt (WDW) patch, the codimension-0 region
bounded by future- and past-directed light sheets emanating from $\Sigma$
on the boundary at the maximal radius $r_m$. Then the complexity is
$C_2=V_{WDW}/G\ell^2$ where $\ell$ is again an arbitrary length scale
\cite{arXiv:1610.02038}.
In the AdS$_5\times S^5$ Poincar\'e patch (\ref{ads5s5}), the future-directed
light sheet is given by $t_+(r)=L^2/r-L^2/r_m$ (and past-directed light sheet
by $t_-=-t_+$), and the complexity comes out to
\beq{ads5cmplx2} \C_2=\frac{L^5\Omega_5}{6G_{10}\ell^2}\frac{r_m^3}{L}
=\frac{1}{6G_5}\left(\frac L\ell\right)^2 \left(\frac{r_m}{L}\right)^3
\ .\eeq
The complexity has the same divergence with $r_m$ as the CV complexity,
and both have the same scaling with AdS radius if the arbitrary length is
chosen as $\ell=L$.

\subsubsection{Parameter dependence}\label{s:cv20scaling}
Since the KS background is static, the CV2.0 complexity is
\beq{cv2element}
\C_2 = \frac{2\OmT}{96G_{10}\ell^2} \int_0^{\tau_m} d\tau\, t_+(\tau,\tau_m)
h(\tau)^{1/2}\sinh^2\tau
\eeq
where the additional factor of $h(\tau)^{-1/4}$ comes from $\sqrt{|g_{tt}|}$,
and the factor of $2t_+(\tau)$ follows from the integral along $t$. Since 
\beq{tplus}
\frac{dt_+}{d\tau}=-\sqrt{\frac{h(\tau)\epsilon^{4/3}}{6K(\tau)^2}} ,\eeq
we see that $t_+=2^{2/3} (g_s M\ap)\epsilon^{-2/3}j_+(\tau,\tau_m)/\sqrt{6}$, where
$j_+(\tau,\tau_m)$ is the dimensionless function defined below. In total,
$\C_2=(g_s M\ap\epsilon)^2\OmT J_2/24\sqrt{6} G_{10}\ell^2$; 
$J_2$ is a dimensionless
form of integral over $\tau$ from (\ref{cv2element}).
This has the same dependence on the confinement scale, $\C_2\propto \epsilon^2$,
as the $\C_V$ complexity.

\subsubsection{Divergence structure}\label{s:cv20divergence}

\begin{figure}[t]\centering
\includegraphics[width=\columnwidth]{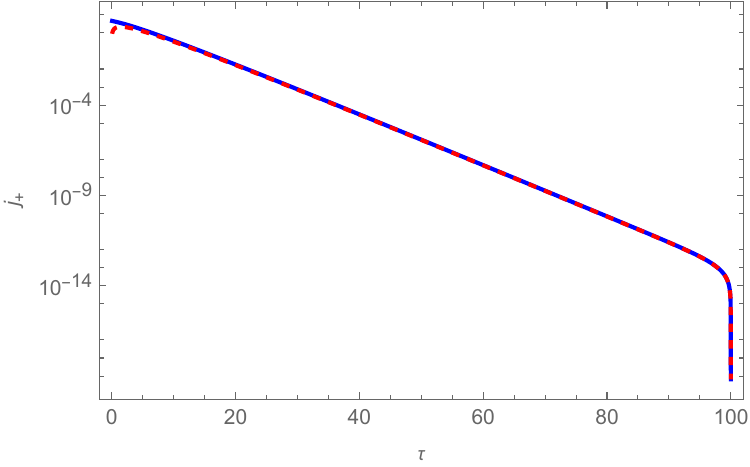}
\caption{\label{f:lightsheet}The dimensionless profile of the WDW patch
future boundary, comparing the numerical calculation [solid blue, 
Eq.~\eqref{lightsheet1}] and leading large $\tau$ approximation [dashed red,
\eqref{lightsheet3}].}
\end{figure}

To find the divergence of the CV2.0 complexity, we need the dimensionless
form of the light sheet 
\beq{lightsheet1}
j_+(\tau,\tau_m) = \int_\tau^{\tau_m} d\eta \frac{I(\eta)^{1/2}\sinh\eta}{(
\sinh (2\eta)-2\eta)^{1/3}} . \eeq
For large values of $\tau$, this is approximately an incomplete gamma function
\bea
j_+(\tau,\tau_m)\!\!\!&\approx&\!\! \frac{\sqrt 3}{2^{5/6}}\int_\tau^{\tau_m} d\eta
\sqrt{\eta-\frac 14} e^{-\eta/3}\nonumber\\
&=&\!\!\frac{9}{2^{5/6}}e^{-1/12} 
\Gamma\left(\frac 32;\frac \tau 3 -\frac{1}{12},\frac{\tau_m}{3}-\frac{1}{12}
\right)\label{lightsheet2}
\eea
including all terms that are not exponentially suppressed. Dropping also
subleading terms in powers of $\tau$ and $\tau_m$, we have
\beq{lightsheet3}
j_+(\tau,\tau_m)\approx \frac{3^{3/2}}{2^{5/6}}\left(\sqrt\tau e^{-\tau/3}-
\sqrt{\tau_m}e^{-\tau_m/3}\right) .\eeq
This is very close to an exponential decay with a rapid cutoff near $\tau_m$
and behaves like AdS in the $r$ coordinate. As expected from the form of 
the power series, we have verified that the relative difference of 
(\ref{lightsheet3}) and the numerically determined value of
(\ref{lightsheet1}) is approximately $1/\tau$, while 
(\ref{lightsheet1}) and (\ref{lightsheet2}) are very close for $\tau\gtrsim 1$;
Fig.~\ref{f:lightsheet} shows the numerically evaluated and leading forms.

The dimensionless form of the CV2.0 complexity is given by the integral
\beq{cv2Jdivergent}
J_2(\tau_m) = \int_0^{\tau_m}d\tau\, j_+(\tau,\tau_m)I(\tau)^{1/2}\sinh^2\tau ,
\eeq
which we evaluate as a double integral by substituting (\ref{lightsheet1}).
If we exchange the order of integration, the integral
\beq{cv2Jdivergent2}
J_2(\tau_m)\sim \int^{\tau_m} d\eta\int^\eta d\tau
\sqrt{\left(\eta-\frac 14\right)\left(\tau-\frac 14\right)} e^{4\tau/3}e^{-\eta/3} 
\eeq
captures the divergence structure of the complexity. This integral does not
appear to have a closed form in terms of commonly known special functions,
though related convergent integrals for $\tau_m\to\infty$ evaluate to
hypergeometric functions.

\begin{figure}[t]\centering
\includegraphics[width=\columnwidth]{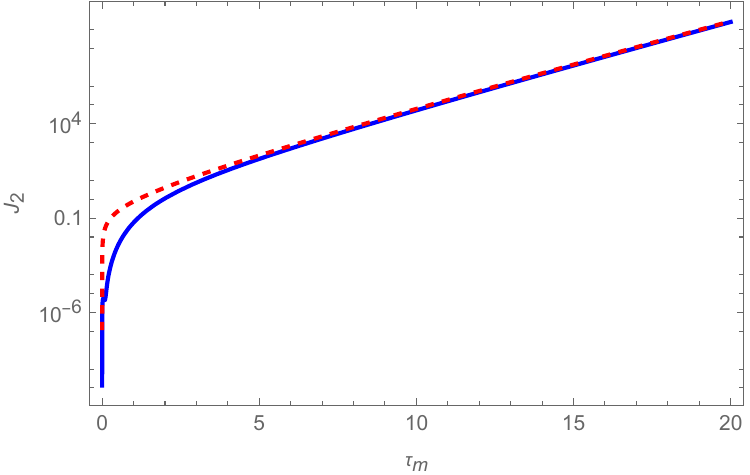}
\caption{\label{f:cv2plot}The dimensionless CV2.0 complexity, evaluated 
numerically [solid blue, Eq.~\eqref{cv2Jdivergent}] 
and as the leading divergence [dashed red, \eqref{cv2Jleading}].}
\end{figure}

If we further ignore terms suppressed by powers of $1/\tau_m$, we find the
leading divergence
\beq{cv2Jleading}
J_2(\tau_m)\approx \frac{9}{32} \tau_m e^{\tau_m} .\eeq
We compare the leading divergence (dashed red) to the numerically evaluated
$J_2(\tau_m)$ (solid blue) in Fig.~\ref{f:cv2plot}. We have confirmed
that the relative difference of the leading divergence (\ref{cv2Jleading}) and 
the numerical value of (\ref{cv2Jdivergent}) scales as $1/\tau_m$ at large
$\tau_m$.

In terms of the radial $r$ coordinate, the complexity is
\beq{cv2leadingr}
\C_2 =\frac{\OmT}{648G_{10}\ell^2} L(r_m)^4r_m^3 .
\eeq
This again agrees with the expectation that the complexity should be the same
as in AdS with a slowly varying curvature scale (up to factors of order
unity). However, it is interesting to note that the divergence differs
from the CV complexity as given in (\ref{cvleadingr}), whereas the divergences
agree for CV and CV2.0 complexity in AdS. Agreement between the two
complexities (up to numerical factors) requires fixing a cutoff and then
choosing the length scale $\ell$ to be the curvature length $L(r_m)$ at
the cutoff.


\section{Action complexity functionals}
In this section, we discuss the CA complexity as the ten-dimensional 
supergravity action, followed by the revised ``CA2.0'' complexity with
additional boundary terms for the supergravity form fields as defined in
\cite{frey}.

\subsection{CA action complexity}\label{s:ca}

\subsubsection{Review}\label{s:careview}
The CA, also known as ``complexity=action,'' proposal takes the on-shell action
of the background evaluated for the WDW patch as a complexity functional
\cite{arXiv:1509.07876,arXiv:1512.04993}.
This action includes boundary terms which take the form of integrals
on the light sheets that form the future and past boundaries of the WDW patch
as well as on the joint at the intersection of those light sheets
\cite{arXiv:1609.00207}. In principle, there can be boundary terms for the form
fields in a 10D supergravity description of the background; however,
agreement between a calculation in AdS$_5\times S^5$ and the usual CA
complexity of AdS$_5$ requires only the bulk action, as we will see below.
Here, we compare the calculations of the CA complexity for AdS$_5\times S^5$
in 5D and 10D since there are some differences.

If we consider AdS$_5$ only, the complexity is the sum of the bulk action
plus a boundary contribution. Since the boundary of the WDW patch consists
of two light sheets emitted from $\Sigma$ on the boundary, the boundary
includes a joint at their intersection on $\Sigma$, which makes its own
contribution. The boundary term includes a logarithmic counterterm necessary
for reparametrization invariance which includes an 
arbitrary length scale $\ell$ (often chosen as the AdS scale $L$). 
The well-known result is $\C_A = \ln(3\ell/L) (r_m/L)^3/4\pi^2 G_5$.

In the 10D type IIB supergravity description,the complexity is
still $C_A=(S_{bulk}+S_{bdry}+S_{joint})/\pi$, though there are some subtleties.
In the bulk action, there is no cosmological constant, which instead
follows from the internal curvature and form fluxes. Furthermore, since the
5-form is self-dual, the commonly used manifestly diffeomorphism invariant 
action is actually a pseudoaction. Instead, we need an action that contains
only independent degrees of freedom, i.e., only half the components of $\t F_5$.
We choose to keep the components $\F_5$ with no legs along the time direction.
If $C_4$ are the components of the 4-form potential also with no legs along
$t$, the bulk action in string frame (and in the absence of free branes) is
\bea
S_{bulk}&=& \frac{1}{2\kappa^2} \int_{WDW} d^{10}x \sqrt{-g} \left(e^{-2\phi}R+
4(\del\phi)^2\right)\nonumber\\
&&-\frac{1}{2\kappa^2} \int_{WDW}\left(\frac 12 e^{-2\phi}
H_3\w\star H_3+\frac 12 \t F_1\w\star\t F_1\right.\nonumber\\
&&\left.+\frac 12\t F_3\w\star\t F_3
+\frac 12\F_5\w\star\F_5-\frac 12 C_4\w\t F_3\w H_3\right.\nonumber\\
&&\left. +\frac 12\F_5\w C_2\w H_3-\frac 12\F_5\w
d_t C_4\right)\label{Sbulk}
\eea
where $d_t=dt\del_t$. In AdS$_5\times S^5$ and the KS background, the
dilaton $\phi=\ln g_s$ is constant, so we take 
$16\pi G_{10}\equiv 2\kappa^2 g_s^2$. In this case, the boundary and joint 
terms are the higher-dimensional analogs of the usual terms in AdS:
\bea
S_{bdry}\!\!&=&\!\!-\frac{1}{8\pi G_{10}}\!
\int_{\del^+}\!\! d\lambda\, d^{8}\sigma\! \sqrt{\gamma}\left(\!\kappa_+\! +\!
\Theta_+\ln|\ell\Theta_+|\vphantom{\frac 12}\!\right)\nonumber\\
&&\!\!+\frac{1}{8\pi G_{10}}\!
\int_{\del^-}\!\! d\lambda\, d^{8}\sigma\! \sqrt{\gamma}\left(\!\kappa_-\!+\!
\Theta_-\ln|\ell\Theta_-|\vphantom{\frac 12}\!\right) ,\ \ \ \
\ \label{Slight sheet}\\
S_{joint}&=&-\frac{1}{8\pi G_{10}}\int_{\del^+\cap\del^-} d^{8}\sigma \sqrt{\gamma} 
\ln\left|k_+\cdot k_-/2\right| .\label{Sjoint}
\eea
Here, $\del^+$ and $\del^-$ indicate the future and past boundaries with 
lightlike parameter $\lambda$ (of dimension length) 
increasing toward the future in both cases. On each light sheet, the vector
$k_\pm=dx/d\lambda$ is the lightlike tangent with $\kappa_\pm$ defined by
$k_\pm^\mu\Del_\mu k_\pm^\nu=\kappa_\pm k_\pm^\nu$. 
The induced metric on each fixed-$\lambda$
spatial slice is $\gamma_{ab}$ (along the $\sigma^a$ coordinates given by 
$\vec x$ and the angular directions), and $\Theta_\pm$ is the logarithmic 
derivative of $\sqrt\gamma$ with respect to $\lambda$.

On AdS$_5\times S^5$, the boundary and joint terms evaluate in essentially
the same fashion as in AdS$_5$ alone. On the other hand, the bulk term
receives no contribution from the Ricci scalar, which cancels between the
positively curved sphere and negatively curved AdS, or from any of the
flux terms except the integral of $-\F_5\star\F_5/2$, which comes to 
$-8/L^2$ integrated over the WDW patch. In the end, we find 
$\C_A=\Omega_5 \ln(3\ell/L) L^2 r_m^3/4\pi^2 G_{10}$,\footnote{One of us
(Frey) originally calculated this result in collaboration with N.~Agarwal 
in separate work \cite{agarwalfrey}.} 
which agrees with the
pure AdS$_5$ result under dimensional reduction. Because of this agreement,
we do not include boundary terms for $\F_5$ in the action, and we also omit
them for $H_3$ and $\t F_3$ for consistency.

\subsubsection{Parameter dependence}\label{s:cascaling}

We consider the action complexity term by term, starting with the bulk 
contribution. In the KS background, none of the forms $C_4$, $\F_5$, $C_2$,
$\t F_3$, or $H_3$ have any legs along the $x^\mu$ directions, so all the
Chern-Simons terms (the last line of (\ref{Sbulk})) vanish. Further, with
constant dilaton, the Einstein equation gives 
$R=(g_s^2|\t F_3|^2+|H_3|^2)/4$,\footnote{We take the convention that 
$|F_p|^2 = F_{M_1\cdots M_p}F^{M_1\cdot M_p}/p!$.}
so the bulk action becomes
\begin{widetext}\bea
S_{bulk}&=&-\frac{1}{16\pi G_{10}}\int_{WDW} d^{10}x\sqrt{-g}\left(
\frac 14 |H_3|^2 +\frac{g_s^2}{4}|\t F_3|^2+\frac{g_s^2}{2} |\F_5|^2\right)
\label{Sbulk2}\\
&=&-\frac{\OmT}{3(16)^2\pi G_{10}}\int d^3\vec x \int_0^{\tau_m}d\tau\, 
t_+(\tau,\tau_m) h(\tau)^{1/2}\sinh^2\tau\left(
\frac 14 |H_3|^2 +\frac{g_s^2}{4}|\t F_3|^2+\frac{g_s^2}{2} |\F_5|^2\right) .
\nonumber\eea\end{widetext}
We can simplify further by noting that the Hodge duality relation between
3-forms implies $|H_3|^2=g_s^2|\t F_3|^2$. Then direct calculation gives us
$g_s^2|\t F_3|^2=3j_3/(g_s M\ap)$ and $g_s^2|\F_5|^2=3j_5/(g_sM\ap)$, where
$j_3$ and $j_5$ are dimensionless functions of $\tau$. We end up with
\begin{widetext}
\beq{Sbulk3} S_{bulk}=
-\frac{g_sM\ap\epsilon^2\OmT}{2(16)^2\pi G_{10}}\sqrt{\frac 23}
\left(\int d^3\vec x\right) \int_0^{\tau_m}d\tau\, j_+(\tau,\tau_m)I(\tau)^{1/2}
\sinh^2(\tau)\left(\vphantom{\frac 12}j_3(\tau)+j_5(\tau)\right) .\eeq
\end{widetext}

The overall scaling of the boundary and joint terms 
(\ref{Slight sheet}) and (\ref{Sjoint}) comes from the spatial slice metric
\beq{slice}
\sqrt{\gamma} = \frac{g_s M\ap \epsilon^2}{16\sqrt 6}\left[\sin\theta_1
\sin\theta_2\right]\sqrt{I(\tau)}\sinh\tau\left(\sinh(2\tau)-2\tau\right)^{1/3}
.\eeq
This also leads to a scaling $\C_A\propto (g_sM\ap)\epsilon^2\OmT$.
However, there is also a logarithmic dependence on $(g_sM\ap)$,
which appears in the joint term through
\beq{jointscale} k_+\cdot k_- = -\left(\frac{1}{\alpha_+\alpha_-\ell^2}\right)
\left[h(\tau)^{-1/2}\left(\frac{dt_+}{d\tau}\right)^2 +
\frac{h(\tau)^{1/2}\epsilon^{4/3}}{6K(\tau)^2}\right] ,\eeq
where we define the lightlike parameter 
$\lambda = \pm\alpha_\pm \ell (\tau_m-\tau)$ on the future (past) light sheet
(note that $k_\pm^\tau =\mp 1/\ell\alpha_\pm$, 
$k_\pm^t=k_\pm^\tau dt_\pm/d\tau=-(1/\ell\alpha_\pm)dt_+/d\tau$).
We have introduced the length scale $\ell$ to give 
$\lambda$ the correct dimensionality; $\alpha_\pm$ are additional arbitrary
constants. From the scaling of the warp factor and $t_+$, we find
$ k_+\cdot k_- = -(g_sM\ap/\ell^2)(j_{joint}/\alpha_+\alpha_-)$, where 
$j_{joint}$ is a dimensionless function of $\tau_m$. 
Once we convert the light sheet actions to integrals over $\tau$,
$\alpha_\pm$ will cancel between the $\ln|\Theta|$ and joint terms.
This leaves $\ln(g_s M\ap/\ell^2)$ as a further parameter dependence.

\subsubsection{Divergence structure}\label{s:cadivergence}

As is typical for the action complexity, finding the divergence structure
$J_A$ is more complicated than for volume complexities.

We begin with the bulk action term. After some simplification, the 
dimensionless flux functions are
\begin{widetext}\bea
j_3(\tau) &=&\frac{1}{4I(\tau)^{3/2}\sinh^6(\tau)}\left[\cosh(4\tau)-24\tau
\sinh(2\tau)+8(1+\tau^2)\cosh(2\tau)+16\tau^2-9\vphantom{\frac 12}\right] ,\quad
\label{Q3}\\
j_5(\tau)&=&\frac{1}{I(\tau)^{5/2}\sinh^6(\tau)}
\left(\tau\coth\tau-1\vphantom{\frac 12}\right)^2\left(\sinh(2\tau)-2\tau
\vphantom{\frac 12}\right)^{4/3} ,\label{Q5}
\eea\end{widetext}
and we define the dimensionless bulk action integral
\beq{caJbulk}J_{bulk}(\tau_m) = \int_0^{\tau_m} d\tau\, j_+(\tau,\tau_m)
\sqrt{I(\tau)}\sinh^2\tau\left( j_3(\tau)+j_5(\tau)\right) .
\eeq
To evaluate this numerically, we substitute (\ref{lightsheet1}) for $j_+$
and calculate $J_{bulk}$ as a double integral. We can also write the
divergent part as
\bea
J_{bulk}(\tau_m)&\sim& 6\sqrt{\frac 23}\int^{\tau_m} d\tau\, 
\frac{2\tau^2/3-\tau/3+5/12}{(\tau-1/4)^2}\nonumber\\
&&\times e^{4\tau/3}\left(\sqrt\tau e^{-\tau/3}
-\sqrt{\tau_m} e^{-\tau_m/3}\right)\nonumber\\
&\approx& \sqrt{\frac 23}\sqrt{\tau_m}e^{\tau_m} .\label{caJbulk2}\eea
It is possible to rewrite this form in terms of $\gamma_1$ or $\gamma^*$ 
functions as in (\ref{tricomi});\footnote{or rather lower incomplete 
versions of those due to divergences at the lower limit in some cases}
however, we keep only the leading divergence as given
because the light sheet action divergence does not have a general closed form.

In the action on the light sheets $\del^\pm$, we can define
$\kappa_\pm\equiv k_\pm^\tau\bar\kappa$ and $\Theta_\pm\equiv k_\pm^\tau\bar\Theta$.
Then 
\bea
S_{bdry}&=&\frac{1}{4\pi G_{10}} \int d^8\sigma\int_0^{\tau_m} d\tau\sqrt\gamma
\left(\bar\kappa+\bar\Theta\ln|\bar\Theta|\right.\nonumber\\
&&\left. +\frac 12\bar\Theta
\ln|k_+^\tau k_-^\tau\ell^2|\right) .\label{Sbdry2}
\eea
Since $k_\pm^\tau$ are constant and $\sqrt\gamma\bar\Theta=\del_\tau\sqrt\gamma$,
the last term is supported only on the joint. Therefore,
\begin{widetext}
\beq{Slsj}
S_{bdry}+S_{joint} = 
\frac{1}{4\pi G_{10}} \int d^8\sigma\int_0^{\tau_m} d\tau\sqrt\gamma
\left(\bar\kappa+\bar\Theta\ln|\bar\Theta|\right)
-\frac{1}{8\pi G_{10}}\int d^8\sigma\sqrt\gamma
\ln\left(\frac{g_sM\ap}{\ell^2}\left|j_{joint}(\tau_m)\right|\right) ,\eeq
where
\bea
\bar\kappa &=&-\frac{1}{K(\tau)}\frac{dK(\tau)}{d\tau} =
\frac{\cosh(2\tau)-6\tau\coth(\tau)+5}{3(\sinh(2\tau)-2\tau)} ,
\nonumber\\
\bar\Theta &=& 
\frac{2I(\tau)(3\coth\tau(\sinh(2\tau)-2\tau)+2(\cosh(2\tau)-1))
-3\csch\tau(\tau\coth\tau-1)(\sinh(2\tau)-2\tau)^{4/3}}{6I(\tau)
(\sinh(2\tau)-2\tau)} ,\nonumber \\
j_{joint} &=& \frac{1}{2g_sM\ap}
\left[h(\tau_m)^{-1/2}\left(\frac{dt_+}{d\tau}\right)^2 +
\frac{h(\tau_m)^{1/2}\epsilon^{4/3}}{6K(\tau_m)^2}\right] =
\frac{\sqrt{I(\tau_m)}\sinh^2\tau_m}{3(\sinh(2\tau_m)-2\tau_m)^{2/3}} .
\label{lsjdefs}\eea
The dimensionless actions are given by $J_{bdry}(\tau_m)+J_{joint}(\tau_m)$,
where
\beq{caJbdry}
J_{bdry}(\tau_m)=\int_0^{\tau_m} d\tau\, \sqrt{I(\tau)}\sinh\tau\left(\sinh(2\tau)
-2\tau\right)^{1/3} \left(\bar\kappa+\bar\Theta\ln\bar\Theta\right) \eeq
and
\beq{caJjoint}
J_{joint}(\tau_m) = \sqrt{I(\tau_m)}\sinh\tau_m\left(\sinh(2\tau_m)
-2\tau_m\right)^{1/3}\ln\left[\left(\frac{g_sM\ap}{\ell^2}\right)
\frac{\sqrt{I(\tau_m)}\sinh^2(\tau_m)}{3(\sinh(2\tau_m)-2\tau_m)^{2/3}}\right]
\eeq
The divergences are
\beq{caJbdry2}
J_{bdry}(\tau_m)\sim \frac 12\sqrt{\frac 32} \int^{\tau_m} d\tau
\sqrt{\tau-\frac 14}e^\tau\left[\frac 13+\left(\frac{4\tau+1}{4\tau-1}\right)
\ln\left(\frac{4\tau+1}{4\tau-1}\right)\right]\approx
\frac{1}{2\sqrt 6} \sqrt{\tau_m}e^{\tau_m}
\eeq
and
\beq{caJjoint2}
J_{joint}(\tau_m)\sim\frac 12 \sqrt{\frac 32} \sqrt{\tau_m-\frac 14}e^{\tau_m}
\ln\left[\frac{1}{2\sqrt{6}} \left(\frac{g_sM\ap}{\ell^2}\right)
\sqrt{\tau_m-\frac 14}\right]\approx
\frac 12 \sqrt{\frac 32} \sqrt{\tau_m}e^{\tau_m}
\ln\left[\frac{1}{2\sqrt{6}} \left(\frac{g_sM\ap}{\ell^2}\right)
\sqrt{\tau_m}\right] .
\eeq\end{widetext}
We again keep only the leading divergence. Note that the expansion terms
($\bar\Theta\ln\bar\Theta$) give only subleading divergences.

\begin{figure}[t]
\centering
\begin{subfigure}[t]{\columnwidth}
\includegraphics[width=\textwidth]{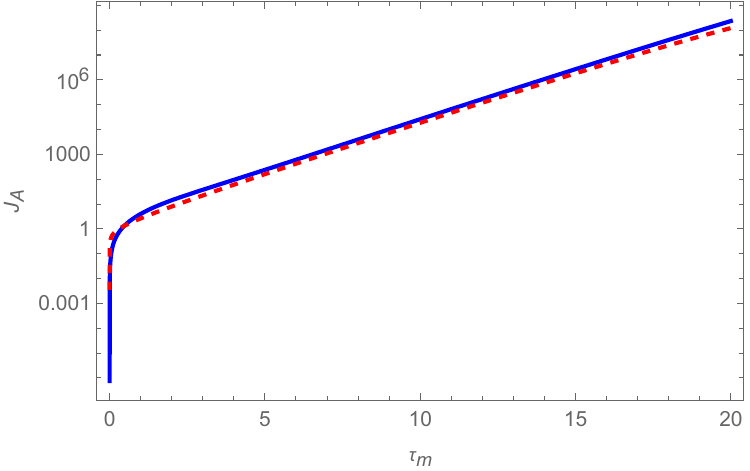}
\end{subfigure}
\begin{subfigure}[t]{\columnwidth}
\includegraphics[width=\textwidth]{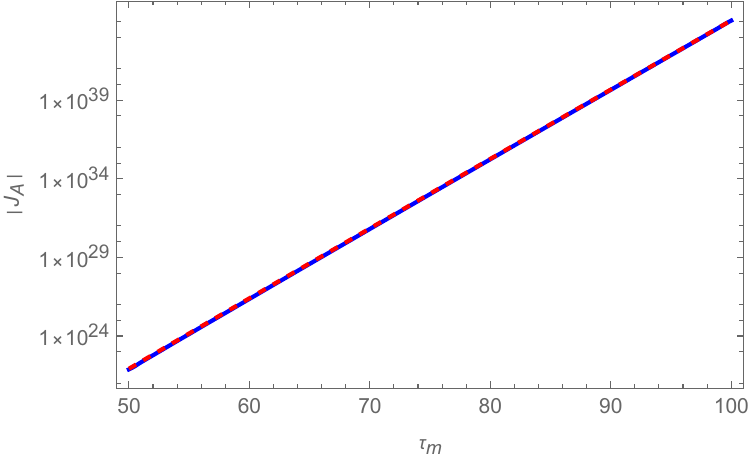}
\end{subfigure}
\caption{\label{f:caplots}The dimensionless CA complexity, evaluated 
numerically [solid blue, from Eqs.~(\ref{caJbulk}), (\ref{Slsj}), 
(\ref{lsjdefs}), (\ref{caJbdry}), (\ref{caJjoint}), and (\ref{cacmplx})] 
and as the leading divergence [dashed red, \eqref{cadivergent}].
To plot, we choose $\ell^2=g_sM\ap$. Note that the complexity becomes
negative at large cutoff, so we show the absolute value.}
\end{figure}

With these definitions, the complexity is
\bea
\C_A &=& \frac{1}{8\pi^2 G_{10}}\frac{g_s M\ap\epsilon^2\OmT}{16\sqrt 6}
\left(-\frac 12 J_{bulk}(\tau_m)\right.\nonumber\\
&&\left.+2J_{bdry}(\tau_m)-J_{joint}(\tau_m)\vphantom{\frac 12}\right)
\nonumber\\
&\equiv&
\frac{1}{8\pi^2 G_{10}}\frac{g_s M\ap\epsilon^2\OmT}{16\sqrt 6}J_A(\tau_m)
 . \label{cacmplx}\eea
Taking the leading divergence, the contributions from the bulk and boundary
cancel to give
\beq{cadivergent}
J_A(\tau_m)\approx -\frac 12\sqrt{\frac 32} \sqrt{\tau_m}e^{\tau_m}
\ln\left[\frac{1}{2\sqrt{6}} \left(\frac{g_sM\ap}{\ell^2}\right)
\sqrt{\tau_m}\right] .\eeq
We show $J_A$ and its leading divergence in Fig.~\ref{f:caplots}. Note
that the sign of $J_A$ changes as $\tau_m$ increases for any fixed value of
the arbitrary length $\ell$. At large $\tau_m$, the relative difference scales
as $1/\tau_m$ as expected.

In terms of the standard radial coordinate $r$, the leading divergence is
\beq{cadivr}
\C_A=\frac{1}{4\pi^2 G_{10}} \frac{\OmT}{108} L(r_m)^2 r_m^3
\ln\left(\frac{3\ell}{L(r_m)}\right) ,\eeq
which is again the same functional form as for AdS$_5\times S^5$ but with
a slowly varying curvature scale. The unusual feature is that the complexity
changes sign with changing cutoff $r_m$ if the arbitrary scale $\ell$ is
fixed. However, from the point of view of evaluating complexity of excited
states of KS (for example, black holes), we should work with a fixed cutoff,
and we are free to choose $\ell$ based on that cutoff [such as $\ell=L(r_m)$].

\subsection{CA2.0 action complexity}\label{s:ca2}
\subsubsection{Review}\label{s:ca2review}

As discussed in \cite{frey}, the gauge transformation of the Chern-Simons
terms in the last line of the bulk action (\ref{Sbulk}) leads to nontrivial
terms on the boundary of the WDW patch. While these vanish on the KS background,
the CA complexity is therefore not gauge invariant on a general background.
As a result, \cite{frey} proposed including additional boundary terms for the
form fields that cancel this gauge variation on shell. There is an infinite
family of boundary terms that gives a gauge-invariant complexity functional;
\cite{frey} chose a set of boundary terms that respect the $SL(2,Z)$ duality
and named the resulting complexity CA2.0, which we denote $C_{\mathbb{A}}$.
On shell, it is possible to write those boundary terms as a contribution to
the bulk integrand, so $C_\mathbb{A} = (S'_{bulk}+S_{bdry}+S_{joint})/\pi$,
where $S_{bdry}$ and $S_{joint}$ are given by (\ref{Slight sheet}) and 
(\ref{Sjoint}), respectively, and
\bea
S'_{bulk}&=& \frac{1}{2\kappa^2} \int_{WDW} d^{10}x \sqrt{-g} \left(e^{-2\phi}R+
4(\del\phi)^2\right)\nonumber\\
&&-\frac{1}{2\kappa^2} \int_{WDW}\left(\frac 14 e^{-2\phi}
H_3\w\star H_3+\frac 12 \t F_1\w\star\t F_1\right.\nonumber\\
&&\left. +\frac 14\t F_3\w\star\t F_3
\right) .\label{Sbulk4}\eea
For AdS$_5\times S^5$ and the KS background, the curvature and modified 
form field strength kinetic terms cancel, so $S'_{bulk}=0$. 
As a result, $C_\mathbb{A}$ is the sum of only the light sheet and joint terms.
For AdS$_5\times S^5$, we have 
$\C_\mathbb{A}=\Omega_5[\ln(3\ell/L)+1/3]L^2r_m^3/4\pi^2G_{10}$.

In the sense that $S'_{bulk}$ vanishes, the CA2.0 complexity presented here is
an extreme, and other gauge-invariant choices for the form boundary terms
should be intermediate between CA2.0 and the standard CA complexity.

\subsubsection{Parameter dependence}\label{s:ca2parameters}
Because the CA2.0 complexity is composed of a subset of the terms from
the CA complexity, the parametric dependence is the same. That is,
$\C_\mathbb{A}\propto (g_s M\alpha')\epsilon^2\Omega_T$ with an additional
logarithmic dependence on $g_sM\alpha'/\ell^2$. 

\subsubsection{Divergence structure}\label{s:ca2divergence}

Following the discussion above, the CA2.0 complexity is 
\bea
\C_\mathbb{A} &=& \frac{1}{8\pi^2 G_{10}} 
\frac{g_s M\alpha'\epsilon^2\Omega_T}{16\sqrt 6} J_\mathbb{A}(\tau_m)\ ,
\nonumber\\
J_\mathbb{A}(\tau_m) &\equiv& 2 J_{bdry}(\tau_m) -J_{joint}(\tau_m)\ ,
\label{ca2cmplx}\eea
where $J_{bdry}$ and $J_{joint}$ are defined in (\ref{caJbdry}) and 
(\ref{caJjoint}). The leading divergences are therefore
\bea
J_\mathbb{A}(\tau_m) &\approx& \frac{1}{\sqrt 6} \sqrt{\tau_m} e^{\tau_m} \left\{
1-\frac 32 \ln\left[ \frac{1}{2\sqrt 6} \left(\frac{g_sM\alpha'}{\ell^2}
\right)\sqrt{\tau_m}\right]\right\}\nonumber\\
&\approx& J_A(\tau_m) .\label{ca2div1}\eea
These terms have the same exponential and power law divergence, though the 
logarithmic term dominates at very large $\tau_m$, so the leading divergence
is the same as for the CA complexity. At large enough $\tau_m$, the CA2.0
complexity follows the right panel of Fig.~\ref{f:caplots}. 

For fixed $\ell$, there is however
an important difference, which is the value $\tau_m^*$ of the cutoff for which 
the complexity changes sign. Because of the additional constant term in the
brackets of (\ref{ca2div1}), $\tau_m^*$ is considerably larger for CA2.0 than
for CA complexity. In fact, $\tau_m^*$ is exponentially sensitive to terms
outside the logarithm, so even the subleading divergences (suppressed above)
shift $\tau_m^*$ noticeably.

In terms of the radial coordinate $r$, (\ref{ca2div1}) gives
\beq{ca2div2}
\C_\mathbb{A} = \frac{1}{4\pi^2 G_{10}} \frac{\OmT}{108} L(r_m)^2 r_m^3
\left[\frac 13 + \ln\left(\frac{3\ell}{L(r_m)}\right)\right] 
\eeq
for the leading divergence.

\section{Flux complexity functionals}
In ten-dimensional supergravity, the flux provides a (local) length scale
due to its dimensionality, much like the curvature. As a result, \cite{frey}
defined two ``flux complexity'' functionals CF and CF2.0 in parallel to the
two volume complexities using the flux to replace the arbitrary length scale.

\subsection{CF flux complexity}\label{s:cf}
\subsubsection{Review}\label{s:cfreview}

The CF complexity (denoted $C_F$) \cite{frey} is the integral
\beq{cfdef1}
C_F \equiv \frac{1}{G_{10}} \int_\B d^9x \sqrt{|g|} \sqrt{\G} , 
\eeq
where $\B$ is the spatial slice that optimizes the integral and
\beq{fluxcombo}
\G\equiv \frac{g_s^2}{16} |\F_5|^2 +\frac{a}{16}\left( |H_3|^2+g_s^2|\t F_3|^2
\right)\ .
\eeq
In AdS$_5\times S^5$, $g_s^2|\F_5|^2=16/L^2$ is constant, so 
$C_F=C_V$ for AdS$_5\times S^5$ with the given normalization
(with $\ell=L$ for CV), and $a$ is an arbitrary constant. 

Compared to 
\cite{frey}, we have set an additional term in $\G$ (which is constant in KS
and AdS$_5\times S^5$) to zero, and we have taken into account a simplification
due to the fact that these flux components have no legs in the time 
direction.

\subsubsection{Parameter dependence}\label{s:cfparameters}

In the notation of Sec.~\ref{s:cascaling}, the flux complexity density is 
\bea
\C_F &=& \frac{\sqrt 6 g_s M\alpha' \epsilon^2\Omega_T}{384 G_{10}}
\int_0^{\tau_m} d\tau\, I(\tau)^{3/4}\sinh^2(\tau )\left( j_5(\tau)+2a j_3(\tau)
\right)^{1/2}\nonumber\\
&\equiv&  \frac{\sqrt 6 g_s M\alpha' \epsilon^2\Omega_T}{384 G_{10}}
J_F(\tau_m) .\label{cfscaling}
\eea
For a fixed $\tau_m$, this shows a different scaling with $g_s M\alpha'$ than
the CV complexity but is instead similar to the action complexities.

\subsubsection{Divergence structure}\label{s:cfdivergence}

\begin{figure}[t]\centering
\includegraphics[width=\columnwidth]{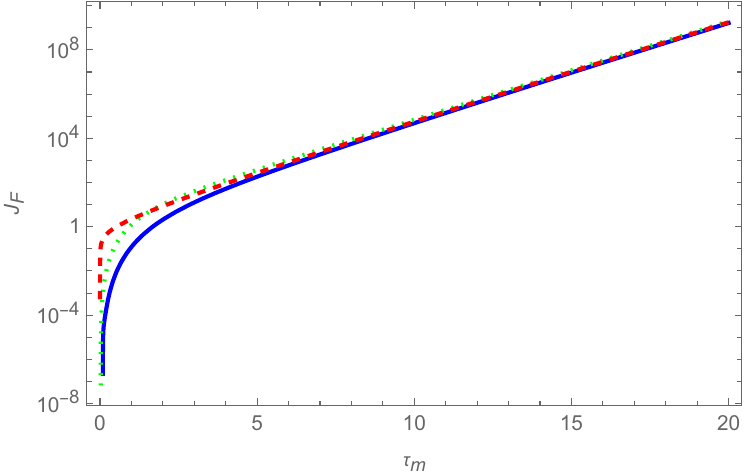}
\caption{\label{f:cfplot} The dimensionless CF complexity $J_F$ as a function
of cutoff $\tau_m$. Compares the numerical calculation for $a=0$ (solid blue) 
and $a=5$ (dotted green) [both from \eqref{cfscaling}] 
to the leading divergence [dashed red, \eqref{cfdiv1}].}
\end{figure}

The dimensionless complexity has the divergence 
\beq{cfdiv1}
J_F(\tau_m) \sim \int^{\tau_m}d\tau\,
\sqrt{a+ \frac 23\frac{(\tau-1)^2}{\tau-1/4} }\,  e^\tau
\approx \sqrt{\frac 23} \sqrt{\tau_m}e^{\tau_m} .
\eeq
Notice that the leading divergence, given after the $\approx$ sign,
is independent of the free parameter $a$ because $j_5/j_3\approx 2\tau/3$.
Therefore, the leading divergence is given entirely by the 5-form flux
rather than the 3-forms; however,
the 3-forms contribute to subleading divergences. 

For illustrative purposes, we have calculated $J_F(\tau_m)$ for both
$a=0$ and $a=5$, which we show in Fig.~\ref{f:cfplot} along with the 
leading divergence. We have verified that the relative difference of the
three curves drops as $1/\tau_m$ at large values of the cutoff.

While CF equals the CV complexity in AdS$_5\times S^5$ when the arbitrary
scale $\ell=L$, the AdS radius, the leading divergence (\ref{cfdiv1})
is softer than the corresponding divergence for CV complexity in KS.
Specifically, if the cutoff is written in terms of the radial coordinate $r$,
the divergence is
\beq{cfdiv2}
\C_F = \frac{\Omega_T}{(18)^2G_{10}} L(r_m)^2 r_m^3 .\eeq
This matches the CV result (\ref{cvleadingr}) precisely when the arbitrary
length is chosen to match the curvature scale at the cutoff $\ell=L(r_m)$.
However, the subleading divergences and finite parts are presumably 
different.

\subsection{CF2.0 flux complexity}\label{s:cf2}
\subsubsection{Review}\label{s:cf2review}

From \cite{frey}, the CF2.0 complexity (indicated with $\mathbb{F}$) 
is given by the integral
\beq{cf20}
C_{\mathbb{F}} \equiv \frac{1}{G_{10}} \int_{WDW} d^{10}x \sqrt{-g}\, \mathcal{G}
\eeq
over the WDW patch with $\mathcal{G}$ as in (\ref{fluxcombo}).
Like the CF complexity equals the CV complexity with the arbitrary length
scale $\ell$ assigned to the AdS scale $L$, this CF2.0 complexity is the 
same as the CV2.0 complexity with $\ell=L$. Specifically, the complexity 
density is $\C_{\mathbb{F}} = \Omega_5 L^2 r_m^3/6G_{10}$.

\subsubsection{Parameter dependence}\label{s:cf2parameters}

Following our previous discussions, the CF2.0 complexity is 
\bea
\C_{\mathbb{F}} &=& \frac{g_sM\alpha' \epsilon^2 \Omega_T}{128\sqrt 6 G_{10}}
J_{\mathbb{F}}(\tau_m)\ ,\label{cf20params}\\ 
J_{\mathbb{F}}(\tau_m) &=& \int_0^{\tau_m} d\tau\, j_+(\tau,\tau_m) \sqrt{I(\tau)}
\sinh^2(\tau)\left( j_5(\tau)+2a j_3(\tau)\right)\ .\nonumber
\eea
Like the CF complexity, the scaling with $g_s M\alpha'$ is the same as in
the action complexities as opposed to the volume complexities, even though
the flux complexities mimic the volume complexities. The reason for the
difference is the absence of an arbitrary length scale in the overall prefactor
in the flux complexities. Even though setting the overall normalization of the
flux complexity is akin to specifying the length scale of the volume 
complexity, it is dimensional analysis that sets the scaling with
$g_s M\alpha'$.

\subsubsection{Divergence structure}\label{s:cf2divergence}

The dimensionless complexity functional $J_{\mathbb{F}}(\tau_m)$ reduces to
the functional $J_{bulk}(\tau_m)$ from CA complexity when $a=1/2$.
Like that integral, we evaluate it numerically by writing $j_+(\tau,\tau_m)$
using (\ref{lightsheet1}) and calculating $J_{\mathbb{F}}(\tau_m)$ as a 
double integral.

Analytically, the divergence part of the complexity is
\bea
J_{\mathbb{F}}(\tau_m) &\sim& 4\sqrt{\frac{3}{2}} \int^{\tau_m}d\tau \left(
\sqrt{\tau}e^{-\tau/3}-\sqrt{\tau_m}e^{-\tau_m/3}\right)e^{4\tau/3}\nonumber\\
&&\times\left[\frac{a+2(\tau-1)^2/3(\tau-1/4)}{\tau-1/4}\right] \label{cf2div1}
\eea
with leading term
\beq{cf2div2}
J_{\mathbb{F}}(\tau_m)\approx \frac{2}{\sqrt 6}\sqrt{\tau_m}e^{\tau_m} .\eeq

\begin{figure}[t]\centering
\includegraphics[width=\columnwidth]{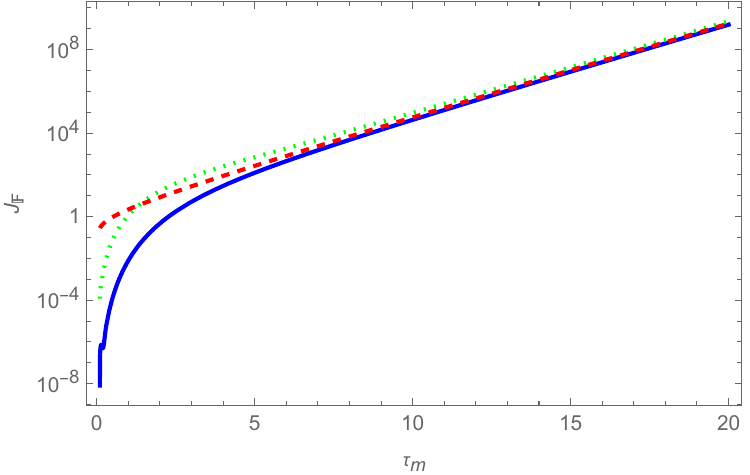}
\caption{\label{f:cf2plot} The dimensionless CF2.0 
complexity $J_{\mathbb{F}}$ as a function
of cutoff $\tau_m$. Compares the numerical calculation for $a=0$ (solid blue) 
and $a=5$ (dotted green) [both Eq.~\eqref{cf20params}] 
to the leading divergence [dashed red, \eqref{cf2div2}].}
\end{figure}

We show the leading CF2.0 divergence in Fig.~\ref{f:cf2plot} along with
numerical calculations of $J_{\mathbb{F}}$ for $a=0$ and $a=5$. For large 
cutoff values $\tau_m$, the difference among these curves decreases as
$1/\tau_m$ as expected. 

In terms of the radial coordinate $r$, the leading divergence is the same
as for the CF complexity, up to the numerical coefficient:
\beq{cf2div3}
\C_{\mathbb{F}} = \frac{\Omega_T}{648 G_{10}} L(r_m)^2 r_m^3 .\eeq
The numerical prefactor is the same as for $\C_2$, but the divergence is
considerably softer for a fixed value of the CV2.0 length scale $\ell$.
The idea that the leading divergence should match based on intuition from
AdS$_5\times X^5$ backgrounds suggests choosing the arbitrary scale to be
the curvature length at the cutoff after fixing the cutoff.
We note, however, that subleading divergences and finite parts will differ
between the CF2.0 and CV2.0 complexities.

\section{Variation on the baryonic branch}\label{s:baryonic}

In this section, we consider the variation of holographic complexity around
the KS background along the baryonic branch. We will work with linear
perturbations and demonstrate that the KS background is an extremum of all
of the complexity functionals considered here (as well as a broad class
of others). 

We begin by describing the baryonic branch linearized around the KS background
as given by \cite{hep-th/0405282} following the discussion in 
\cite{hep-th/0412187}. At the linear level in deformation parameter $q$, 
the warp factor $h(\tau)$ is 
unchanged, as are $\t F_3$ and $\F_5$. In terms of the basis forms $g_i$
of (\ref{gbasis}), the linear deformation of $H_3$ is 
\beq{deltaH}
\delta H_3 = q\chi'(\tau) d\tau\wedge\left(g_2\wedge g_3-g_1\wedge g_4\right) .
\eeq
Also writing the metric in terms of the $g_i$ basis, 
\bea
\delta\t g_{13} &=& \delta\t g_{24} = q\cosh(\tau)K(\tau) Z(\tau)\ ,\nonumber\\
\delta \t g^{13}&=&\delta \t g^{24}= -q\cosh(\tau)\coth^2(\tau)K(\tau) Z(\tau) .
\label{deltametric}\eea
In the above, $\chi'(\tau)$ and $Z(\tau)$ can be written in terms of 
hypertrigonometric functions, but their precise form is not important to us.
Finally, the axiodilaton is unperturbed (and remains constant). We note
that \cite{hep-th/0412187} constructed the baryonic branch backgrounds only 
along a one-parameter curve in the moduli space because other solutions
are related by S duality; see more discussion below.

We can now see that the constituents of the complexity functionals discussed
here have vanishing first-order variation along the baryonic 
branch.\footnote{We have also verified these results by evaluating the
ansatz of \cite{hep-th/0412187} on the linearized baryonic branch around the
KS background.}
First, because $\delta\t g_{mn}$ is traceless and the warp factor is unchanged
at linear order, neither the nine-dimensional volume measure of a constant $t$ 
slice nor the ten-dimensional volume measure (within the WDW patch) changes.
Next, it is straightforward to see by checking indices that
\beq{F3pert}
\delta\left(|\t F_3|^2\right) = h^{-3/2}(\tau)\frac 12 \t F_{mnp}\t F_{qrs} 
\t g^{mq}\t g^{nr} \delta\t g^{ps}=0\ .
\eeq
Similarly, $\delta(|\F_5|^2) =0$.
The square of the Neveu-Schwarz--Neveu-Schwarz 
3-form flux is slightly more complicated:
\bea
\delta\left(|H_3|^2\right) &=& h^{-3/2}(\tau)\frac 12 H_{mnp} H_{qrs}
\t g^{mq}\t g^{nr} \delta\t g^{ps}\label{H3pert}\\ 
&&+ h^{-3/2}(\tau)\frac 13 H_{mnp}\delta H_{qrs}
\t g^{mq}\t g^{nr} \t g^{ps} =0\nonumber
\eea
also by comparing the components of $H_3$ and $\delta H_3$. Additionally, we
can see that $\t F_{mnp}H^{mnp}$ has zero variation. Finally, since the
Ricci scalar is given by a sum of $|\t F_3|^2$ and $|H_3|^2$, it is also
unchanged. (Further, the Ricci tensor is diagonal, so powers of it are also 
unperturbed.)

Taken together, these results mean that the first variation of all the 
complexity functionals discussed in this paper vanishes when evaluated on the
KS background. A large family of more general complexity functionals are 
also extremized on the KS background for the same reason. 
Complexity can be defined in quantum mechanics as the distance of the
state in question from some reference state along an optimal path in state
space \cite{quant-ph/0502070,quant-ph/0603161,quant-ph/0701004}; following
the first law of complexity \cite{Bernamonti:2019zyy,Bernamonti:2020bcf},
we see that the first variation along the baryonic branch is orthogonal
to that optimal path. We also note that our calculation agrees with the
first law of complexity in that the variation of complexity depends only on
the variation of the target state (namely the supergravity background in
our case).

It is worth considering why complexity might be extremized at KS solution
along the baryonic branch; we speculate that there is a rotational symmetry
of the moduli space with a fixed point at the KS solution. In that case,
smoothness (specifically absence of a kink) of complexity as a function of
baryonic modulus requires its derivative to vanish at the KS point.
This rotational symmetry is not immediately obvious in the usual description
of the baryonic branch in the gauge theory, which is parametrized by
a complex variable $\xi$ with $\xi=1$ at the $\mathbb{Z}_2$ symmetric point
\cite{hep-th/0405282}. However, as discussed in \cite{hep-th/0412187},
the supersymmetry of the supergravity dual can be parametrized by a real 
function $\alpha$ and a complex function $\beta$ with the KS background
at $\beta=0$; \cite{hep-th/0412187} (following \cite{hep-th/0405282}) found
the baryonic branch solution for imaginary $\beta$. 
Real and imaginary $\beta$ are related by S duality, so 
derivatives of the complexity in those directions must vanish 
at $\beta=0$ to be compatible with S duality and smoothness.\footnote{We 
speculate that the supergravity $SL(2,\mathbb{R})$ symmetry is related to
rotations in the $\beta$ plane, although we do not know the detailed mapping.} 
Smoothness then implies that the variation
of the complexity must vanish along any direction of the baryonic branch 
at the KS background. If this reasoning holds, 
the extremization of complexity at 
the KS background is a consequence of duality.

\section{Discussion}\label{s:discussion}

We have evaluated the holographic complexity of the Klebanov-Strassler
background with several complexity functionals, including the CA2.0,
CF, and CF2.0 functionals proposed recently in \cite{frey} based on 
considerations from the 10D supergravity. 
The KS background is an
interesting test case for holographic complexity in both 10D and 
nonconformal gravity backgrounds, and we compare the complexity in KS to
that of AdS$_5\times S^5$ (as evaluated in 10D variables).
At large radius, KS is approximately AdS$_5\times T^{1,1}$ with a 
logarithmically varying curvature length $L(r)$; substituting this
value $L(r_m)$ at the cutoff for the AdS scale transforms the complexity of 
AdS$_5\times T^{1,1}$ into the leading divergence of the complexity of KS
(up to numerical factors).
Specifically, this leading divergence is $f(L(r_m))r_m^3$, where $f(L)$
is a power or a power times a logarithm, depending on the complexity 
functional.

In fact, if we insist that the AdS complexity determines the leading 
divergence, rewriting that divergence in terms of the dimensionless radial
coordinate $\tau$ controls the dependence on the gauge theory parameters
$g_s$, $M$, and $\epsilon$. Most importantly, the universal $r_m^3$ 
factor is $\epsilon^2\exp(\tau_m)$ (times numerical factors of order unity), 
and $\epsilon^2$ is the cube of the confinement
length. This is the same scaling as the complexity of formation for 
AdS solitons, which may suggest a universal role for confinement in
complexity. Meanwhile, the dependence on $g_s$ and $M$ follows from the
functional dependence on $L(r_m)$. An interesting point is that the dependences
on $\epsilon$ and with $g_sM$ both rely on how we fix parameters:
is the cutoff radius a fixed $\tau_m$ or $r_m$, and do we fix the 10D
Newton constant or the 5D one in an effective theory?

However, the complexity of KS is not only the leading term, which diverges
faster than the AdS complexity due to the growth of $L(r_m)$. There is 
also a series of divergent terms of the form $\exp(\tau_m)\sum_p \tau_m^{n-p}$
(times $\ln\tau_m$ for some complexities), as well as a finite contribution.
As a result, the dependence on $g_s$, $M$, and $\epsilon$ is simplest if
we define the cutoff radius as a fixed $\tau_m$. Furthermore, even the
leading divergence differs among complexity functionals if arbitrary
lengths $\ell$ in the functional definitions are held fixed. For example,
if the CV, CV2.0, and CA functionals are to have parametrically similar values,
we should choose $\ell$ for all three as the UV length scale $L(r_m)$
(times a constant). We can ask if this is a hint toward associating the
arbitrary scale $\ell$ with a UV scale.

The full complexity of the KS ground state, which we determine numerically
for several functionals, is necessary to define the complexity of formation
for excited backgrounds like thermal (black brane) states (the complexity of
formation is the difference between the excited and ground state complexities). 
Given that there
may be no closed form even for all the divergences in terms of commonly known
special functions, it may be useful to make a modified definition for 
complexity of formation in asymptotically KS backgrounds. On the other hand,
it may be possible to define a complexity of formation by subtracting the
integrands when writing the complexities as integrals over $\tau$. An 
interesting important question is whether exciting the KS background can 
change the subleading divergences, which would seem to make it impossible to
define a finite complexity of formation in the standard way.

We make a point of comparison to other work.
Recently, \cite{Fatemiabhari:2024aua} proposed a complexity functional
for 10D backgrounds involving an overall conformal factor in the metric.
In the KS solution, the complexity density of \cite{Fatemiabhari:2024aua}
is\footnote{We thank C.~Nu\~nez for providing this formula.} 
\beq{nunez}\C\propto \epsilon^4\int d\tau\, h(\tau)\sinh^2(\tau),\eeq
which has a different scaling with $\epsilon$ and a leading divergence
of a different form [i.e., not $f(L(r_m))r_m^3$]. These differences highlight
the need to understand the relation between different complexity functionals.

Somewhat more speculatively, our work may be helpful in understanding the
holographic complexity of de Sitter (dS) spacetime \cite{arXiv:2202.10684} 
in the context of string theory. Specifically, an important class of
dS compactifications \cite{hep-th/0301240} includes KS or similar warped
throats in the compactification geometry.

Finally, as a separate point, because the KS gauge theory has a moduli
space (the baryonic branch), the variation of the complexity along that
moduli space is a physically meaningful quantity. We have shown for a 
broad family of complexity functionals, including all six functionals
considered here, that the holograpahic complexity is extremized on the 
KS background. That is, the first variation of the complexity along the
baryonic branch, including
all divergences and finite contributions, vanishes on the KS background.

\acknowledgments
ARF would like to thank N.~Agarwal for conversations and collaboration on
separate but related work and C.~Nu\~nez for interesting discussion. 
This work was supported by the
Natural Sciences and Engineering Research Council of Canada Discovery Grant
program, grant 2020-00054. MPG was additionally supported
by the Natural Sciences and Engineering Research Council of Canada USRA
program. PS was additionally supported by the Mitacs Globalink program.

\section*{Data Availability}
The data that support the findings of this article are openly available
\cite{ksdata}.


\bibliography{kscomplexity}

\end{document}